\documentclass[aps,twocolumn,nofootinbib, preprintnumbers, superscriptaddress]{revtex4}

\usepackage{amsmath, amssymb, slashed, braket, bm}
\usepackage{graphicx}
\usepackage{epstopdf}
\usepackage{float,appendix}
\usepackage[colorlinks=true,
            linkcolor=blue,
            urlcolor=blue,
            citecolor=green,
            bookmarks=true,
            bookmarksnumbered=true,
            breaklinks=true,
            pdfpagemode=FullScreen,
            pdfstartview=FitBH]{hyperref}
\usepackage{esint}
\usepackage[normalem]{ulem}
\usepackage{siunitx}
\usepackage{multirow}
\usepackage{xcolor}

\usepackage{xcolor}
\definecolor{gesfpurple}{rgb}{0.47,0.19,0.42}

\definecolor{gesflanse}{rgb}{0.00,0.50,0.50}

\definecolor{gesfblue}{rgb}{0.08,0.42,0.76}

\definecolor{gesfred}{rgb}{1,0,0}

\definecolor{gesfwhite}{rgb}{1,1,1}

\definecolor{gesfblack}{rgb}{0,0,0}

\newcommand{\prlsection}[2]{{\it\textbf{#1}{#2}}---}

\def\MPL{M_\text{Pl}}

\begin{document}

\title{The LUX-ZEPLIN Event as Hyperfine Spectroscopy of Composite Dark Matter}

\author{Jie Sheng}
\email{jie.sheng@ipmu.jp}
\affiliation{
Kavli IPMU (WPI), UTIAS, University of Tokyo, Kashiwa, 277-8583, Japan}

\author{Kairui Zhang}
\email{kzhang25@ou.edu}
\affiliation{Homer L. Dodge Department of Physics and Astronomy, University of Oklahoma, 73019, USA}

\begin{abstract}
    We propose that the $248\,\mathrm{keV}$ nuclear-recoil candidate reported by LUX--ZEPLIN can arise from the endothermic hyperfine excitation of composite dark matter. Rather than requiring two distinct dark-sector particles, the inelasticity of scattering arises from a hyperfine transition between the pseudoscalar and vector states of a single dark hadron. In a PQ-augmented dark QCD framework, the parity-conserving vacuum suppresses dangerous elastic scattering, while axion emission depletes the relic excited-state population. Endothermic excitation near the Galactic kinematic edge then naturally produces a localized high-energy recoil peak. We find that sub-TeV-scale dark matter with a hyperfine splitting of a few hundred keV can yield an order-one event rate in the LZ high-recoil window. This interpretation offers a possible form of dark-matter hyperfine spectroscopy, with complementary tests from dark-photon searches and future recoil-spectrum measurements.
\end{abstract}

\maketitle

\noindent
\prlsection{Introduction}{.}
The particle nature of dark matter (DM) remains unknown~\cite{Cirelli:2024ssz}. Liquid-xenon time-projection chambers currently provide leading sensitivity to a broad class of weak-scale DM interactions~\cite{Billard:2021uyg}. The LUX--ZEPLIN (LZ) Collaboration has recently extended its nuclear-recoil analysis to approximately $270$~keV and reported a nuclear-recoil candidate with energy
\begin{equation}\label{eq:LZ_ER}
    E_R = 248\pm23\,(\mathrm{stat})\pm23\,(\mathrm{sys})~\mathrm{keV}
\end{equation}
in a $2.84$ tonne-year exposure. Among the tested signal models, the maximum local significance is $3.4\sigma$, reduced to a global significance of $2.6\sigma$ by the look-elsewhere effect~\cite{LZ:2026axp}. A DM interpretation is difficult to reconcile with conventional elastic scattering, whose spectrum favors low-energy recoils and is suppressed at high energies by the halo distribution and nuclear form factor.

Endothermic scattering, $\chi_1 N\to\chi_2 N$ with $\Delta=m_{\chi_2}-m_{\chi_1}>0$, changes this expectation~\cite{Tucker-Smith:2001myb} and has emerged as a leading interpretation of this event~\cite{Su:2026rwz,Freese:2026sga,Fan:2026kxx,Wu:2026nhi,DiMauro:2026ldr,Smirnov:2026aqk,McCabe:2026crm,Langhoff:2026ujr,2609.15985,Yin:2026jnn,Du:2026guj,Bisal:2026khf,Visinelli:2026kgt,Kotlarski:2026pep,Ahmed:2026qjg,Lee:2026xxh,Nomura:2026qyq,Wang:2026ytg,Bandyopadhyay:2026gjw,Borah:2026zwf,Lee:2026jxl,Okada:2026upm,Borah:2026ris,2609.15742,Yamashita:2026ump,Lee:2026wof,Okada:2026eol,Du:2026lpa,Zhu:2026dag,Das:2026uyy,Yuan:2026djt,Qi:2026vyp,Kumar:2026lgi,Cabo-Almeida:2026uqw,Barman:2026omh,Xing:2026civ,Asadi:2026lz09107,Ge:2026xax}. The excitation threshold selects the high-velocity halo tail, while scattering near threshold confines recoils to a narrow energy interval, producing a localized high-energy feature. This interpretation requires a splitting of a few hundred keV, much smaller than the DM mass, and a strongly suppressed elastic channel to avoid an accompanying low-energy signal. Exothermic scattering and DM absorption offer alternative explanations by converting dark-sector mass energy into nuclear recoil energy~\cite{Graham:2010ca,deLima:2026shq,Baer:2026fpy,Fan:2026hzw,2609.15714,Lou:2026idn,Dror:2019onn,Dror:2019dib,Dror:2020czw,Ge:2022ius,PandaX:2022ood}.

Higgsino DM provides one example~\cite{Fan:2026kxx,Freese:2026sga,Wu:2026nhi,Yin:2026jnn,Du:2026guj,Bisal:2026khf,Visinelli:2026kgt,Kotlarski:2026pep,Ahmed:2026qjg,Lee:2026xxh}. When the electroweak gauginos are much heavier than the Higgsinos, the two lightest neutralinos are Higgsino-like and nearly degenerate~\cite{Baer:2024tfo,Baer:2026zra}. Besides the fact that such a SUSY scenario can be challenging in collider searches~\cite{Baer:2022qqr,Baer:2022smj,Baer:2023yxk,Baer:2023olq,Baer:2025zqt,Baer:2025uzx,Zhang:2026eoc,Baer:2026zra} and hard to reconcile with naturalness~\cite{Baer:2024hpl,Baer:2024hgq,Baer:2023uwo,Baer:2026zra,Baer:2024fgd}, its conventional thermal Higgsino interpretation is also in tension with solar-neutrino searches\footnote{Departures from thermal freeze-out permit much heavier Higgsino DM, for which reduced solar capture can allow an interpretation of the LZ event~\cite{Feldstein:2013uha,Langhoff:2026ujr}.}: Higgsinos captured in the Sun through the same inelastic nuclear scattering responsible for explaining the LZ event can subsequently annihilate into Standard Model (SM) weak gauge bosons, producing secondary high-energy neutrinos strongly constrained by IceCube and Super-Kamiokande~\cite{Pospelov:2026ewn,Bose:2026ndd}. Similar bounds in principle apply to other models with efficient solar capture and sufficiently rapid annihilation into neutrino-producing final states. An asymmetric relic can evade these bounds through its suppressed antiparticle abundance~\cite{Zurek:2013wia,Nagata:2026pbj}.

Rather than attributing the endothermic energy splitting to the mass difference between two distinct dark-sector particles, in this work, we propose that the required splitting arises from an internal hyperfine excitation of composite DM. As is known in SM QCD, mesons containing charm or bottom quarks exhibit nearly degenerate spin partners, such as $B$--$B^*$ and $\eta_c$--$J/\psi$, with the same quark flavor content but different spin configurations. When the heavy-quark mass $m_h$ greatly exceeds the confinement scale $\Lambda_d$, the leading strong interactions are independent of its spin. Subleading hyperfine interactions break this spin symmetry and split the spin partners~\cite{Neubert:1993mb}. This has motivated composite inelastic DM (CiDM), originally proposed to explain the DAMA/LIBRA anomaly~\cite{Alves:2009nf,SpierMoreiraAlves:2010err}. In such a framework, DM is a heavy--light dark meson with a pseudoscalar ground state $\pi_d$ and a nearby vector excitation $\rho_d$, whose hyperfine splitting determines the threshold energy for $\pi_dN\to\rho_dN$.

However, applying a vanilla CiDM model to the LZ event requires several additional ingredients. With the heavy dark-quark mass far above the confinement scale, efficient annihilation in the dark strong sector tends to leave the symmetric relic underabundant. A primordial asymmetry can supply the required relic abundance, but must be protected against washout through confinement~\cite{Alves:2009nf,SpierMoreiraAlves:2010err}. The stability of $\pi_d$ must also be protected despite the presence of lighter dark quark–antiquark mesons. An axially coupled mediator suppresses the leading elastic channel provided the dark strong sector conserves parity. A generic physical dark vacuum angle, however, violates $P$ and $CP$ and can induce an unacceptably large elastic signal. Finally, a cosmologically long-lived vector excitation $\rho_d$ population can produce low-energy exothermic recoils~\cite{Finkbeiner:2009mi}.

To explain the LZ event within CiDM while addressing all these challenges, we introduce a dark Peccei--Quinn (PQ) symmetry~\cite{PhysRevLett.38.1440,PhysRevD.16.1791}. The PQ charge assignments yield an accidental heavy–light flavor symmetry that stabilizes $\pi_d$ and protects the primordial asymmetry. The construction also provides a UV origin for the dark-quark mass hierarchy. The dark axion $a_d$ dynamically relaxes the physical vacuum angle to a parity-conserving minimum, suppressing the otherwise dangerous elastic scattering. It also enables the fast decay $\rho_d\to\pi_d a_d$, depleting the excited population and suppressing exothermic scattering. We identify parameters for which the endothermic hyperfine transition $\pi_d N\to\rho_d N$ can account for the LZ event.

\vspace{0.3cm}
\noindent
\prlsection{Model}{.}
We consider an $SU(N)_d\times U(1)_{A'}$ gauge theory with four left-handed Weyl fermions $\chi_{h,l}$ and $\chi_{h,l}^c$. A dark Higgs $\Phi$ breaks $U(1)_{A'}$, while a gauge-singlet scalar $S$ breaks the global PQ symmetry~\cite{PhysRevD.16.1791,PhysRevLett.38.1440,PhysRevLett.40.279,PhysRevLett.40.223,Sheng:2025sou}. Table~\ref{tab:charges} lists their charges. The fermion content is vectorlike under $SU(N)_d$, and the $U(1)_{A'}$ gauge anomalies cancel.

\begin{table}[t]
\centering
\renewcommand{\arraystretch}{1.13}
\setlength{\tabcolsep}{3.8pt}
\begin{tabular}{c|cccccc}
    \hline
    \hline
     & $\chi_h$ & $\chi_h^c$ & $\chi_l$ & $\chi_l^c$ & $\Phi$ & $S$ \\
    \hline
    $SU(N)_d$ & $\Box$ & $\overline{\Box}$ & $\Box$ & $\overline{\Box}$ & $\mathbf{1}$ & $\mathbf{1}$ \\
    $U(1)_{A'}$ & $1$ & $1$ & $-1$ & $-1$ & $-2$ & $0$ \\
    $U(1)_{\rm PQ}$ & $0$ & $0$ & $1$ & $2$ & $0$ & $-3$ \\
    \hline
    \hline
\end{tabular}
\caption{Charge assignments of the dark sector. All fermions are left-handed Weyl fields.}
\label{tab:charges}
\end{table}

The gauge symmetry forbids bare flavor-diagonal dark-quark masses, while the PQ symmetry forbids both bare flavor-mixing masses and the renormalizable light-quark Yukawa coupling. The leading mass-generating interactions are
\begin{equation}
     -\mathcal L_Y
        \supset
             y_h\Phi\chi_h\chi_h^c
             +\frac{y_l}{\Lambda_{\rm UV}}S\Phi^\dagger\chi_l\chi_l^c
             +\mathrm{h.c.},
     \label{eq:new_yukawa}
\end{equation}
where $\Lambda_{\rm UV}$ is a UV cutoff above the PQ-breaking scale. After spontaneous breaking of $U(1)_{A'}$ and $U(1)_{\rm PQ}$, in unitary gauge, one can expand
\begin{equation}
    \Phi=\frac{1}{\sqrt2} (v_\Phi + s_\Phi),
    \quad
    S=\frac{f_{a_d}+s}{\sqrt2}e^{-ia_d/f_{a_d}},
    \label{eq:new_vevs}
\end{equation}
with the radial modes $s_\Phi$ and $s$ decoupled from the low-energy theory. The Weyl fields form two Dirac fermions, $\Psi_{h,l}=(\chi,\chi^{c\dagger})^T_{h,l}$, with masses
\begin{equation}
     m_h=\frac{y_hv_\Phi}{\sqrt2},
     \quad
     m_l=\frac{y_lf_{a_d}v_\Phi}{2\Lambda_{\rm UV}}.
     \label{eq:constituent_masses}
\end{equation}
The PQ symmetry thus generates the heavy–light quark mass hierarchy. Breaking $U(1)_{A'}$ also gives the dark photon a mass $m_{A'}=2g_Av_\Phi$, where $g_A$ is its gauge coupling.

The dark photon $A'$ and dark axion $a_d$ couple to the dark matter currents through
\begin{equation}
    -\mathcal L_d\supset
    g_A J_d^\mu A'_\mu
    +\frac{\partial_\mu a_d}{3f_{a_d}}J_{\rm PQ}^\mu.
    \label{eq:def_interactions}
\end{equation}
The $SU(N)_d$ interaction confines at the scale $\Lambda_d$, binding the dark quarks into hadrons. Above this scale,
\begin{align}
    J_d^\mu 
        &= \bar\Psi_h\gamma^\mu\gamma_5\Psi_h -\bar\Psi_l\gamma^\mu\gamma_5\Psi_l, \label{eq:gauge_current}\\
    J_{\rm PQ}^\mu
        &= -\frac12\bar\Psi_l\gamma^\mu\Psi_l -\frac32\bar\Psi_l\gamma^\mu\gamma_5\Psi_l.
 \label{eq:pq_current}
\end{align}

Below $\Lambda_d$, the appropriate degrees of freedom are dark hadrons. The lightest heavy–light hadrons are the pseudoscalar $\pi_d$ and vector $\rho_d$,
\begin{align}
     \pi_d&\sim\bar\Psi_l\gamma_5\Psi_h,
     & J^P&=0^-,
     \nonumber\\
     \rho_d^\mu&\sim\bar\Psi_l\gamma^\mu\Psi_h,
     & J^P&=1^-.
\end{align}
For $m_h\gg\Lambda_d$, heavy-quark spin symmetry makes these states degenerate at leading order, with a common mass $m_h + \mathcal O(\Lambda_d)$. The degeneracy is broken by the dark chromomagnetic spin--spin interaction~\cite{Neubert:1993mb},
\begin{equation}
     \delta m
        \simeq\frac{\kappa_{\rm hf}\Lambda_d^2}{m_h}\langle\mathbf S_h\cdot\mathbf S_l\rangle +\mathcal O\left(\frac{\Lambda_d^3}{m_h^2}\right),
\end{equation}
where $\kappa_{\rm hf}$ is an $\mathcal O(1)$ hadronic coefficient. For total spin $\mathbf J=\mathbf S_h+\mathbf S_l$,
\begin{equation}
     \langle\mathbf S_h\cdot\mathbf S_l\rangle
     =\frac12\left[J(J+1)-\frac34-\frac34\right]
     =\begin{cases}-3/4,&J=0 \\+1/4,&J=1\end{cases}
\end{equation}
leaving a hyperfine splitting
\begin{equation}
     \Delta\equiv m_{\rho_d}-m_{\pi_d}
     \simeq\kappa_{\rm hf}\frac{\Lambda_d^2}{m_h}.
     \label{eq:hfsplit}
\end{equation}
To explain the LZ anomaly, one needs
\[
    m_{\pi_d} \simeq m_h \sim 0.4-4~\mathrm{TeV}, \quad \Delta \sim 300~\mathrm{keV}.
\]
For $\kappa_{\rm hf}\sim \mathcal{O}(1)$, this implies $\Lambda_d\sim\mathcal{O}(100)$ MeV. Together with the requirement to fulfill the desired hierarchy $m_l \lesssim \Lambda_d \ll m_h$, and assuming both Yukawa couplings areof $\mathcal{O}(1)$, the suppression in Eq.~\eqref{eq:new_yukawa} is at least $\Lambda_{\rm UV}/f_{a_d} \gtrsim 10^{3}$.

The gauge and PQ charge assignments in Table~\ref{tab:charges} also yield accidental vector flavor symmetries $U(1)_h\times U(1)_l$, under which
\begin{equation}
    \Psi_h\to e^{i\alpha_h}\Psi_h,
    \quad
    \Psi_l\to e^{i\alpha_l}\Psi_l,
\end{equation}
with both scalar fields neutral. The corresponding flavor numbers $Q_h$ and $Q_l$ are separately conserved. These symmetries are free of dark gauge anomalies and remain unbroken by the scalar vacuum expectation values and confinement\footnote{For similar discussions of accidental symmetry-protected DM stability or symmetry quality issues, see Refs.~\cite{Baer:2025srs,Baer:2025oid,Sheng:2025sou,Sheng:2026aro}.}.

As the lightest state with $(Q_h,Q_l)=(1,-1)$, the meson $\pi_d\sim (\Psi_h\bar \Psi_l)$ is stabilized by the accidental flavor conservation despite the presence of lighter light-quark hadrons. The same symmetries preserve the primordial flavor asymmetries and are respected by the $\rho_d\to\pi_d a_d$ transition. Operators violating $Q_h$ or $Q_l$ first appear at dimension six for $N=2$ and at dimension seven or higher for $N>2$.

Matching the currents in Eq.~\eqref{eq:def_interactions} onto the $\pi_d$--$\rho_d$ matrix elements gives the effective meson currents relevant to the LZ transition and the decay transition (Appendix~\ref{app:matching}),
\begin{align}
     J_{d,\rho\pi}^\mu&=
     -\frac{C_2^A}{\Lambda_d}\,
     \partial_\nu
     \left(\pi_d^\dagger
     \overleftrightarrow{\partial^\mu}\rho_d^\nu\right)
     +\mathrm{h.c.},
     \label{eq:meson_gauge_current}\\
     J_{\rm PQ,\rho\pi}^\mu&=
     -m_{\pi_d}C_{\rho\pi}\,
     \pi_d^\dagger\rho_d^\mu+\mathrm{h.c.},
     \label{eq:meson_pq_current}
\end{align}
where $C_2^A$ and $C_{\rho\pi}$ are hadronic form factors, taken to be $\sim \mathcal{O}(1)$.

To connect to the LZ event, the dark photon must also couple to the SM. This can be achieved by kinetic mixing with hypercharge~\cite{Fabbrichesi:2020wbt},
\begin{equation}
     -\mathcal L_{\rm mix}\supset
     \frac{\epsilon}{2c_W}F'_{\mu\nu}B^{\mu\nu}.
     \label{eq:darklag}
\end{equation}
Diagonalizing the kinetic and mass terms gives
\begin{equation}
     -\mathcal L_{\rm int}
        \supset A'_\mu\bigg(
     g_AJ_d^\mu+\epsilon eJ_{\rm EM}^\mu
     +\frac{\epsilon e}{c_w^2}
     \frac{m_{A'}^2}{m_Z^2-m_{A'}^2}J_Z^\mu
     \bigg)+\mathcal O(\epsilon^2),
     \label{eq:portal_currents}
\end{equation}
where $c_w=\cos\theta_W$, $s_w=\sin\theta_W$, $J_{\rm EM}^\mu=\sum_f Q_f\bar f\gamma^\mu f$, and $J_Z^\mu=\sum_f\bar f\gamma^\mu(T_f^3P_L-s_w^2Q_f)f$. 

An axial dark-photon coupling alone does not guarantee the absence of elastic scattering. A nonzero dark vacuum angle $\bar\theta_d$ violates $P$ and $CP$ and can induce unwanted elastic scattering in $\langle\pi_d(p')|J_d^\mu|\pi_d(p)\rangle$. The PQ symmetry, however, is anomalous under both $SU(N)_d$ and $U(1)_{A'}$. From Table~\ref{tab:charges},
\begin{equation}
    \mathcal A_{{\rm PQ}[SU(N)_d]^2}=\frac{3}{2}, \quad \mathcal A_{{\rm PQ}[U(1)_{A'}]^2}=3N
\end{equation}
Dark confinement generates an axion potential minimized at $\bar\theta_d+\langle a_d\rangle/f_{a_d}=0$, selecting a parity-conserving vacuum. The domain-wall number is
\begin{equation}
    N_{\rm DW}=\frac{2\mathcal A_{{\rm PQ}[SU(N)_d]^2}}{|Q_{\rm PQ}(S)|}=1
\end{equation}
so the model is free of the domain-wall problem~\cite{PhysRevLett.43.103,Shifman:1980if}. The curvature of this potential determines the axion mass~\cite{GrillidiCortona:2015jxo},
\begin{equation}
    m_{a_d}^2f_{a_d}^2=\chi_d(0),
\end{equation}
where $\chi_d(0)$ is the vacuum topological susceptibility of the dark strong sector. Crucially, since 
\[
    m_l\lesssim\Lambda_d, \quad \chi_d(0)^{1/4}\lesssim\Lambda_d,
\]
one automatically has
\begin{equation}\label{eq:axion_mass_bound}
    m_{a_d}\lesssim
    0.25\,\mathrm{eV}
    \left(\frac{\Lambda_d}{0.5\,\mathrm{GeV}}\right)^2
    \left(\frac{10^9\,\mathrm{GeV}}{f_{a_d}}\right).
\end{equation}
Thus, for $f_{a_d}\gg m_h \gg \Lambda_d$, the axion is much lighter than the hyperfine splitting $\Delta$, allowing $\rho_d\to\pi_d a_d$ decay through the PQ transition current in Eq.~\eqref{eq:meson_pq_current}. Its width is
\begin{equation}
    \Gamma_{\rho_d\to\pi_d a_d}
    \simeq
    \frac{|C_{\rho\pi}|^2\Delta^3}
         {216\pi f_{a_d}^2}.
    \label{eq:rhopiawidth}
\end{equation}
The corresponding lifetime is $\tau_{\rho_d} = \Gamma_{\rho_d\to\pi_d a_d}^{-1}$
\begin{equation}
    \tau_{\rho_d}
    \simeq
    9.57\times10^6\,\mathrm{s}\,
    |C_{\rho\pi}|^{-2}
    \left(\frac{360\,\mathrm{keV}}{\Delta}\right)^3
    \left(\frac{f_{a_d}}{10^9\,\mathrm{GeV}}\right)^2.
    \label{eq:rho_lifetime}
\end{equation}
For $f_{a_d}\sim10^9\,\mathrm{GeV}$ and $|C_{\rho\pi}|\sim1$, this decay depletes the primordial $\rho_d$ population before structure formation, eliminating its exothermic recoil contribution while preserving the total heavy--light meson abundance. Appendix~\ref{app:cosmology} further discusses the cosmological implications.

\vspace{0.3cm}
\noindent
\prlsection{Endothermic scattering}{.}
For the transition $\pi_d N\to\rho_d N$, a nuclear recoil of energy $E_R$ corresponds to a momentum transfer
\begin{equation}
    Q^2\equiv|\mathbf q|^2=2m_NE_R,
    \label{eq:qER}
\end{equation}
where $m_N$ is the nuclear mass. The minimum incident DM speed to produce this recoil is~\cite{Tucker-Smith:2001myb}
\begin{equation}
    v_{\min}(E_R)=
    \frac{m_NE_R/\mu_{\pi_d N}+\Delta}
         {\sqrt{2m_NE_R}},
    \label{eq:vminup}
\end{equation}
with the reduced mass $\mu_{\pi_d N}=m_{\pi_d}m_N/(m_{\pi_d}+m_N)$. Minimizing it gives the excitation threshold,
\begin{equation}
    v_{\rm th}=\sqrt{\frac{2\Delta}{\mu_{\pi_d N}}},
    \quad
    E_R^\star=\frac{\mu_{\pi_d N}}{m_N}\Delta.
    \label{eq:threshold}
\end{equation}
For an incident speed $v\geq v_{\rm th}$, the allowed recoil energies lie between
\begin{equation}
    E_R^\pm(v)=\frac{\mu_{\pi_d N}^2}{2m_N}
    \left(v\pm\sqrt{v^2-v_{\rm th}^2}\right)^2.
    \label{eq:recoilendpoints}
\end{equation}

Adopting a Galactic escape speed $v_{\rm esc}=544\,\mathrm{km/s}$ and an Earth speed $v_{\rm Earth}=254\,\mathrm{km/s}$ relative to the halo, the maximum incident speed is
\begin{equation}
    v_{\max}=v_{\rm esc}+v_{\rm Earth}
    =798\,\mathrm{km/s}.
    \label{eq:vmax}
\end{equation}
The accessible recoil interval is therefore $E_R^-(v_{\max})\leq E_R\leq E_R^+(v_{\max})$. When $v_{\rm th}$ lies just below $v_{\max}$, only the high-velocity tail of the halo can scatter, and both recoil endpoints approach $E_R^\star$, as can be seen from Eq.~\eqref{eq:recoilendpoints}, localizing the signal at nonzero recoil energy.

\begin{figure}[t]
    \centering
    \includegraphics[width=1\columnwidth]{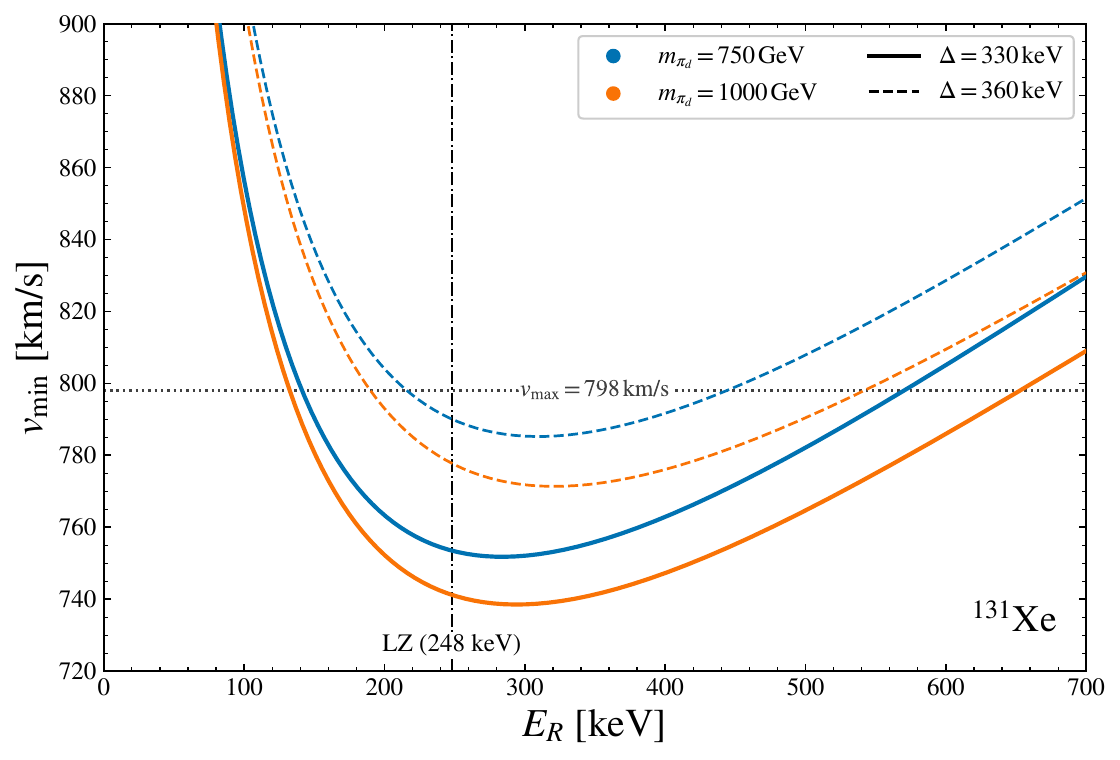}
    \caption{
        Minimum incident speed for endothermic scattering on $^{131}\mathrm{Xe}$. Blue and orange curves correspond to $m_{\pi_d}=750$ and $1000\,\mathrm{GeV}$, respectively; solid and dashed curves show $\Delta=330$ and $360\,\mathrm{keV}$. The horizontal dotted line marks $v_{\max}=798\,\mathrm{km/s}$, below which recoils are kinematically accessible. The vertical dash-dotted line marks the central LZ recoil energy, $E_R=248\,\mathrm{keV}$.
    }
    \label{fig:vmin}
\end{figure}

Fig.~\ref{fig:vmin} illustrates these effects for $^{131}\mathrm{Xe}$. The intersections of each curve with the horizontal $v_{\max}$ line determine the accessible recoil endpoints $E_R^\pm(v_{\max})$. At fixed $m_{\pi_d}$, increasing $\Delta$ shifts the minimum to higher recoil energy and raises the threshold speed $v_{\rm th}$, narrowing the accessible interval. This suppresses low-energy recoils while reducing the available halo population capable of scattering, eventually closing the channel when $v_{\rm th}>v_{\max}$. On the other hand, at fixed $\Delta$, increasing $m_{\pi_d}$ increases the reduced mass, lowers $v_{\min}$ and shifts the minimum slightly upward. The resulting wider recoil interval admits more halo particles, but also allows recoils farther above and below the LZ event energy. The choice of $m_{\pi_d}$ and $\Delta$ therefore balances the localization of the signal against the available halo population: a larger splitting restricts the recoil interval, while a heavier DM mass relaxes the excitation threshold at the cost of broader kinematic support. The actual recoil spectrum within this interval depends on the halo velocity distribution and the scattering response.

The scattering amplitude follows from the dark transition current in Eq.~\eqref{eq:meson_gauge_current} and the nuclear electromagnetic current. In the nonrelativistic limit, the latter is dominated by its temporal component,
\begin{equation}
    \langle N(k')|J_{\rm EM}^0|N(k)\rangle
        \simeq 2m_N Z F_N(Q^2),
    \label{eq:nuclear_em_current}
\end{equation}
where $Z$ is the atomic number and $F_N(Q^2)$ is the nuclear charge form factor,
\begin{equation}
    F_N(Q^2)=
    \frac{\displaystyle\int_0^\infty dr\,
          r^2\rho(r)j_0(Qr)}
         {\displaystyle\int_0^\infty dr\,
          r^2\rho(r)}.
    \label{eq:nuclear_form_factor}
\end{equation}
Here $\rho(r)$ is the nuclear charge density and $j_0(x)=\sin x/x$ is the spherical Bessel function. We adopt the two-parameter Fermi fit to the SCRIT $^{132}\mathrm{Xe}$ measurement~\cite{Tsukada:2017llu},
\begin{equation}
    \rho(r)=\frac{\rho_c}{1+\exp[(r-c)/a]},
    \quad
    c=5.4\,\mathrm{fm},\quad a=0.614\,\mathrm{fm}.
    \label{eq:fermi_density}
\end{equation}
The normalization $\rho_c$ cancels in Eq.~\eqref{eq:nuclear_form_factor}. We use this profile for all natural-xenon isotopes, with each isotope’s mass entering $Q^2$.

Over the narrow momentum-transfer range relevant to the LZ event, we treat the form factor $C_2^A(q^2) \simeq C_2^A$ as an $\mathcal{O}(1)$ constant. Summing over the final vector polarizations gives the leading nonrelativistic squared amplitude,
\begin{equation}
    \overline{|\mathcal M_{\rm scat}|^2}
    \simeq
    \frac{
        32g_A^2\epsilon^2e^2Z^2
        m_{\pi_d}^2m_N^3E_R
    }{
        \Lambda_d^2(Q^2+m_{A'}^2)^2
    }
    |C_2^A|^2|F_N(Q^2)|^2.
    \label{eq:amplitude}
\end{equation}
The differential cross section is then
\begin{equation}
\begin{split}
    \frac{d\sigma_{\rm scat}}{dE_R}
    &=
    \frac{\overline{|\mathcal M_{\rm scat}|^2}}
         {32\pi m_{\pi_d}^2m_Nv^2}
    \\
    &\simeq
    \frac{m_N^2}{m_{\pi_d}\Delta}
    \frac{4\alpha Z^2}{f_{\rm eff}^4}
    \frac{E_R}{v^2}
    |F_N(Q^2)|^2
    \left(1+\frac{Q^2}{m_{A'}^2}\right)^{-2}.
\end{split}
    \label{eq:crossup}
\end{equation}
Here $v$ is the incident DM speed and $\alpha=e^2/(4\pi)$. We have used $\Lambda_d^2\simeq m_{\pi_d}\Delta/\kappa_{\rm hf}$ and defined the effective interaction scale $f_{\rm eff}$ by
\begin{equation}
    \frac{1}{f_{\rm eff}^4}
    \equiv
    \frac{\kappa_{\rm hf}|C_2^A|^2g_A^2\epsilon^2}
         {m_{A'}^4}.
    \label{eq:feff}
\end{equation}

If $\rho_d$ survives on cosmological timescales, the same interaction also mediates exothermic scattering $\rho_d N\to\pi_d N$. For an unpolarized incident $\rho_d$ population, averaging over its spin states gives an additional factor of $1/3$ relative to Eq.~\eqref{eq:crossup}, with the recoil interval determined by exothermic kinematics.

\vspace{0.3cm}
\noindent
\prlsection{Parameter space for the LZ event}{.}
Assuming $\pi_d$ constitutes all local DM, the differential recoil rate per unit detector mass for a single isotope is~\cite{Lewin:1995rx}
\begin{equation}
    \frac{dR_{\rm scat}}{dE_R}
        =
        \frac{\rho_0}{m_{\pi_d}m_N}
        \int_{v_{\min}(E_R)}^{v_{\max}}
        d^3v\,v f(\mathbf v)
        \frac{d\sigma_{\rm scat}}{dE_R},
        \label{eq:rate}
\end{equation}
where $\rho_0 = 0.45\,$GeV/cm$^3$ is the local DM density and $f(\mathbf v)$ is the normalized Maxwellian velocity distribution boosted to the Earth's frame~\cite{Baxter:2021pqo}. Summing over natural-xenon isotopes gives $dR_{\rm scat}^{\rm Xe}/dE_R$. Applying the LZ efficiency $\varepsilon_{\rm LZ}(E_R)$ gives the expected event count for an exposure $\mathcal E=2.84,\mathrm{tonne,yr}$~\cite{LZ:2026axp},
\begin{equation}
    N_{\rm det}
        =
        \mathcal E
        \int_{202\,\mathrm{keV}}^{294\,\mathrm{keV}}
        dE_R\,
        \varepsilon_{\rm LZ}(E_R)
        \frac{dR_{\rm scat}^{\rm Xe}}{dE_R}.
    \label{eq:event_count}
\end{equation}
The integration window corresponds to $248\,\mathrm{keV}\pm1\sigma_{\rm stat} \pm1\sigma_{\rm sys}$ in Eq.~\eqref{eq:LZ_ER}. For each $m_{\pi_d}$ and $\Delta$, we fix $f_{\rm eff}$ by requiring one detected event in this window:
\begin{equation}
    N_{\rm det}(248~\mathrm{keV} \pm 1\sigma_{\rm stat} \pm 1\sigma_{\rm sys})=1.
\end{equation}

\begin{figure}[t]
    \centering
    \includegraphics[width=1\columnwidth]{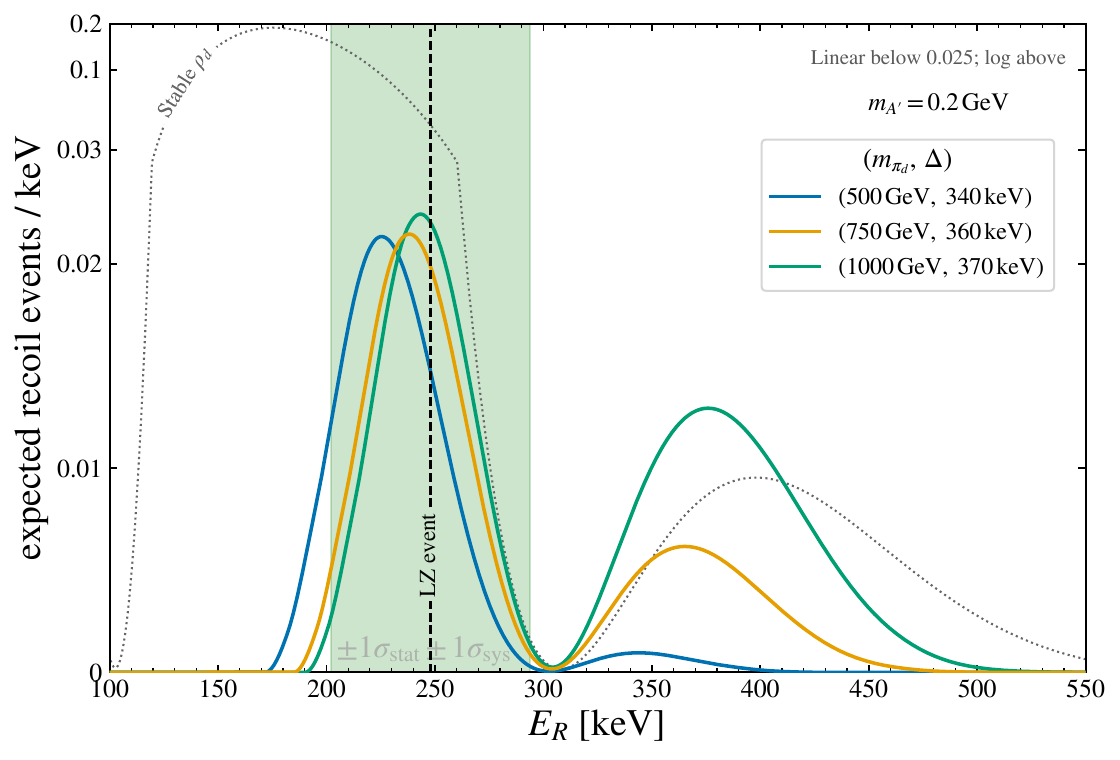}
    \caption{
        Physical recoil spectra before detector efficiency for the benchmarks in Table~\ref{tab:numerical_setup}. Each colored curve is normalized to a single detected (efficiency applied) event in the shaded LZ window. The gray dotted curve shows the would-be exothermic contribution with a representative excited fraction $n_{\rho_d}/n_{\pi_d}\sim10^{-4}$, in the absence of $\rho_d\to\pi_d a_d$ decay.
    }
    \label{fig:spectrum}
\end{figure}

\begin{table}[t]
    \centering
    \begingroup
    \renewcommand{\arraystretch}{1.18}
    \setlength{\tabcolsep}{7pt}
    
    \begin{tabular}{l@{\hspace{10pt}}|@{\hspace{10pt}}ccc}
        \hline\hline
        Parameter & \multicolumn{3}{c}{Value} \\
        \hline
        $m_{\pi_d}$ [GeV]
            & $500$ & $750$ & $1000$ \\
        $\Delta$ [keV]
            & $340$ & $360$ & $370$ \\
        \hline
        $m_{A'}$ [GeV]
            & \multicolumn{3}{c}{$0.2$} \\
        $\rho_0$ [GeV/cm$^3$]
            & \multicolumn{3}{c}{$0.45$} \\
        $v_0$ [km/s]
            & \multicolumn{3}{c}{$238$} \\
        $v_{\rm max}$ [km/s]
            & \multicolumn{3}{c}{$798$} \\
        \hline\hline
    \end{tabular}
    \endgroup
    \caption{
    Benchmark parameters and common inputs for Fig.~\ref{fig:spectrum}. Here $v_0$ is the characteristic speed of the Maxwellian halo distribution in the Galactic frame.
    }
    \label{tab:numerical_setup}
\end{table}

Fig.~\ref{fig:spectrum} shows the benchmark spectra from Table~\ref{tab:numerical_setup}. Each colored curve is, however, normalized to one detected (efficiency applied) event in the shaded window via Eq.~\eqref{eq:event_count}. The pronounced minimum around $300\,\mathrm{keV}$ arises from a diffraction minimum of the xenon nuclear charge form factor, beyond which the nuclear response rises again and produces a secondary high-energy feature. Across these benchmarks, heavier DM produces a more pronounced high-energy tail. Excessive high-energy tails, however, can create tension with the absence of events in the LZ high-energy sideband, as emphasized in recent studies~\cite{Rodd:2026tyn,Dent:2026bji}. The lighter benchmark suppresses this tail, but has more leakage into the low-energy region and a main peak farther below the LZ event energy. At fixed $m_{\pi_d}$, increasing $\Delta$ shifts the peak toward higher recoil energies and suppresses the low-energy leakage. This, however, raises $v_{\rm th}$ toward $v_{\max}$ and reduces the halo population capable of scattering, as illustrated in Fig.~\ref{fig:vmin}, requiring a stronger interaction to maintain the one-event normalization.

The gray curve shows the would-be exothermic contribution for a surviving excited fraction $n_{\rho_d}/n_{\pi_d}\sim10^{-4}$ when $\rho_d\to\pi_da_d$ is absent (Appendix~\ref{app:rho_d}). Even this tiny population produces substantial low-energy recoils because exothermic scattering has no excitation threshold. Axion emission entirely depletes $\rho_d$, eliminating this contribution.

\begin{figure}[t]
\centering
\includegraphics[width=1\columnwidth]{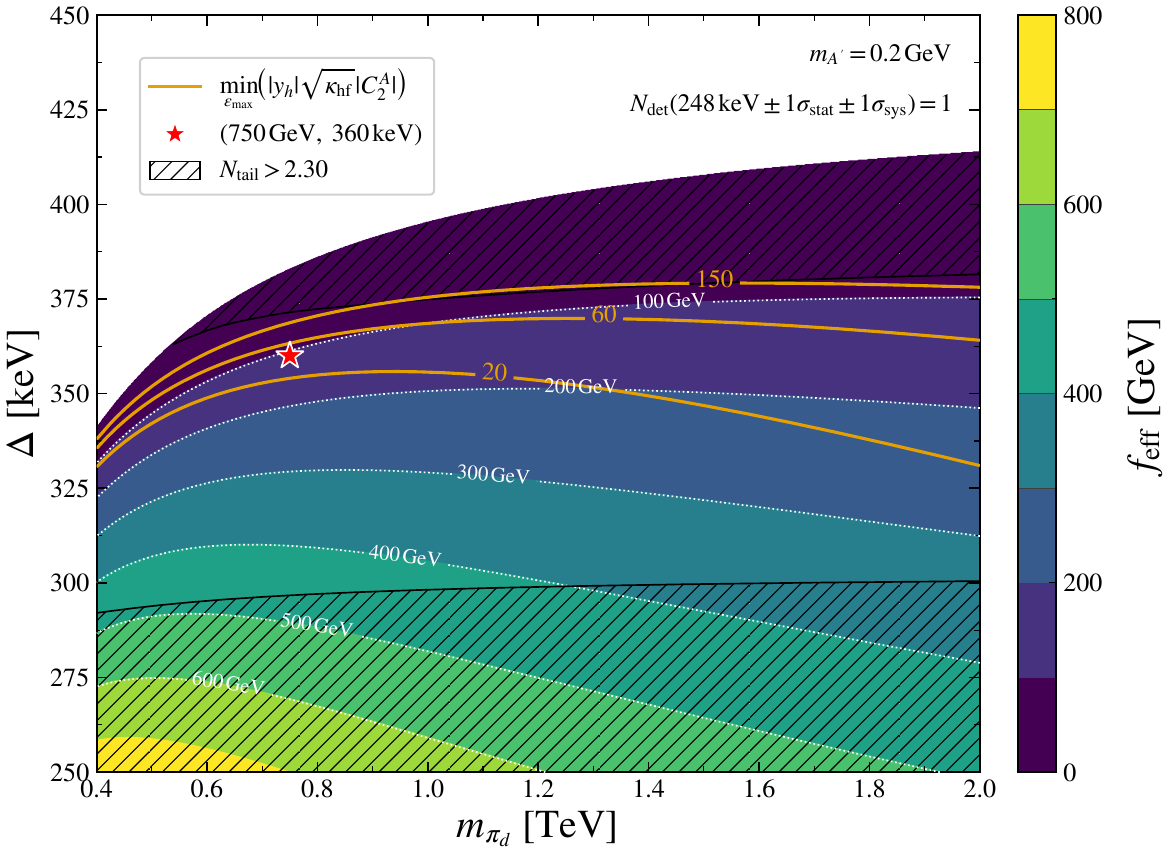}
\caption{
    The color scale and white contours show $f_{\rm eff}$ required for one detected (efficiency applied) event in the LZ window. Orange contours show the minimum $|y_h|\sqrt{\kappa_{\rm hf}}|C_2^A|$ consistent with the visible dark-photon bound [Eq.~\eqref{eq:coupling_bound}]. Black hatching marks $N_{\rm tail}>2.30$ [Eq.~\eqref{eq:tail_count}]. The red star is the benchmark corresponding to the orange spectrum in Fig.~\ref{fig:spectrum}.
}
\label{fig:scan}
\end{figure}

Fig.~\ref{fig:scan} shows the $f_{\rm eff}$ required for one detected event across the $(m_{\pi_d},\Delta)$ plane. Increasing $\Delta$ reduces the halo population contributing to the LZ window, requiring smaller $f_{\rm eff}$ and a stronger interaction. Above the colored region, the window is kinematically inaccessible.

Using $m_{A'}=2g_Av_\Phi$ and $m_{\pi_d}\simeq m_h=|y_h|v_\Phi/\sqrt2$, and Eq.~\eqref{eq:feff}, the kinetic mixing can be expressed as
\begin{equation}
    |\epsilon|
    \simeq
    \frac{2\sqrt2\,m_{\pi_d}m_{A'}}
    {|y_h|\sqrt{\kappa_{\rm hf}}|C_2^A|
     f_{\rm eff}^2}.
    \label{eq:epsilon_from_feff}
\end{equation}
The single-event normalization fixes $f_{\rm eff}$. For a given $m_{A'}$ and $m_{\pi_d}$, the experimental upper limit on kinetic mixing can be translated into a required lower bound on
\begin{equation}
    |y_h|\sqrt{\kappa_{\rm hf}}|C_2^A|
    \gtrsim
    \frac{2\sqrt2\,m_{\pi_d}m_{A'}}
    {\epsilon_{\max}f_{\rm eff}^2}
    \label{eq:coupling_bound}
\end{equation}
For $m_{A'}=0.2\,\mathrm{GeV}<2m_\mu$, the dimuon channel is closed and inapplicable~\cite{LHCb:2019vmc}. On the other hand, the light dark mesons have masses of order $\Lambda_d\simeq \sqrt{m_{\pi_d}\Delta/\kappa_{\rm hf}} \sim \mathcal{O}(100)~\mathrm{MeV}$ for $\mathcal{O}(1)$ $\kappa_{\rm hf}$. Hence, $A'$ is below the dark-hadron decay threshold as well. The dark photon therefore decays predominantly
into $e^+e^-$. We adopt~\cite{BaBar:2014zli}
\[
    \epsilon_{\max} \simeq9.4\times10^{-4}
\]
at this mass. The orange contours show the minimum coupling combination obtained by saturating this bound. Large required values would imply either nonperturbative $y_h$ or hadronic coefficients $\kappa_{\rm hf}$ and $|C_2^A|$ larger than their assumed $\mathcal O(1)$ values.

The
\begin{equation}
    N_{\rm tail}
        = N_{\rm det}(5\,\text{--}202\,\mathrm{keV})+
    N_{\rm phys}(E_R>300\,\mathrm{keV})
\label{eq:tail_count}
\end{equation}
counts the expected tail events, where $N_{\rm phys}$ counts recoils without detector efficiency for the same exposure because LZ~\cite{LZ:2026axp} does not publish the efficiency in this region. The black hatching marks $N_{\rm tail}>2.30$, motivated by the 90\% Poisson upper limit for zero observed events. Since the high-energy contribution is evaluated before detector efficiency, the hatching represents a strict criterion on the recoil spectrum. The lower hatched region is dominated by leakage below the LZ recoil window. At large $\Delta$, the window becomes increasingly difficult to populate, and the stronger interaction required by the normalization enhances the surviving high-energy tail, producing the upper hatched region. The intermediate region balances these two contributions. The red star marks the $(m_{\pi_d},\Delta)=(750\,\mathrm{GeV},360\,\mathrm{keV})$ benchmark, corresponding to the orange curve in Fig.~\ref{fig:spectrum}. Appendix~\ref{app:cosmology} shows that the parameter region explaining the LZ event is also consistent with cosmological constraints.

\vspace{0.3cm}
\noindent
\prlsection{Conclusion and discussion}{.}
The recent LZ high-recoil event is difficult to accommodate with conventional elastic WIMP scattering and has motivated inelastic interpretations involving both endothermic and exothermic scattering. A key question is then what generates the required energy splitting and how low-energy recoil events can be suppressed. While it may arise from distinct dark sector particles with different masses, we point out that it can instead originate from different internal states of the same composite DM~\cite{Bai:2013xga,Kribs:2016cew,Kaplan:2009de,Kaplan:2011yj,Cline:2013pca,Qiu:2023bbp}, for example, the dark hadron, through their binding-energy splitting.

In the PQ-completed axial CiDM framework, a confining dark-QCD sector produces a pseudoscalar ground state $\pi_d$ and a nearby vector excitation $\rho_d$, separated by a hyperfine splitting. The dark axion dynamically relaxes the dark strong-CP phase, suppressing parity-violating elastic scattering, and provides a decay channel that reduces the present excited fraction. The leading transition $\pi_dN\to\rho_dN$ then combines an endothermic threshold with a momentum-dependent transition operator, suppressing conventional low-energy recoils and producing a high-energy peak in the spectrum. One-event normalizations in the high recoil energy window require for sub-TeV to TeV DM masses and splittings of roughly $0.3\,$MeV.

The kinetically mixed dark photon provides a complementary experimental probe. For our benchmark mass $m_{A'}=0.2\,\mathrm{GeV}$, it decays predominantly into $e^+e^-$. Visible dark-photon searches and recoil-spectrum measurements jointly test this interpretation of the LZ event as hyperfine spectroscopy of dark hadrons.

\vspace{0.3cm}
\noindent
\prlsection{Acknowledgements}{.}
The authors thank Prof. Shigeki Matsumoto and Prof. Tsutomu T. Yanagida for useful discussions.
J.~S.\ is supported by the Japan Society for the Promotion of Science (JSPS) as a
part of the JSPS Postdoctoral Program (Standard) with grant number: P25018. K.~Z.\ gratefully acknowledges support from the Avenir Foundation.

\providecommand{\href}[2]{#2}\begingroup\raggedright\endgroup

\appendix
\onecolumngrid
\vspace{0.3cm}
\section{Meson Matrix Element Matching}\label{app:matching}

Starting from the Lagrangian in Eq.~\eqref{eq:def_interactions}
\begin{equation}
    -\mathcal L_d
        \supset
        g_A J_d^\mu A'_\mu
        +\frac{\partial_\mu a_d}{3f_{a_d}}J_{\rm PQ}^\mu,
\end{equation}
we match the constituent currents in Eqs.~\eqref{eq:gauge_current} and \eqref{eq:pq_current}
\begin{equation}
     J_d^\mu 
        = \bar\Psi_h\gamma^\mu\gamma_5\Psi_h -\bar\Psi_l\gamma^\mu\gamma_5\Psi_l
    \quad 
    J_{\rm PQ}^\mu
        = -\frac12\bar\Psi_l\gamma^\mu\Psi_l -\frac32\bar\Psi_l\gamma^\mu\gamma_5\Psi_l.
\end{equation}
onto transition currents between $\pi_d$ and $\rho_d$. Lorentz invariance and parity conservation determine the axial transition elements as
\begin{equation}
\begin{split}
    \langle\pi_d(p)|J_d^\mu|\rho_d(p',\varepsilon)\rangle
    ={}
    m_{\pi_d}C_1^A(q^2)\varepsilon^\mu
    +\frac{C_2^A(q^2)}{\Lambda_d}
    (p+p')^\mu(q\cdot\varepsilon)
    &+\frac{C_3^A(q^2)}{\Lambda_d}
    q^\mu(q\cdot\varepsilon) + \cdots
\end{split}
\label{eq:matching_gauge}
\end{equation}
where $q=p'-p$, $\varepsilon^\mu$ is the $\rho_d$ polarization, and $C_i^A(q^2)$ are dimensionless hadronic form factors, taken to be of $\mathcal O(1)$. The factor $m_{\pi_d}$ reflects the relativistic normalization of the heavy meson states, while $\Lambda_d$ is inserted by dimensional analysis. The ellipsis denotes contributions from higher-dimensional meson operators suppressed by additional powers of $\Lambda_d$.

For nuclear scattering, the $C_3^A$ contribution vanishes upon contraction with the conserved electromagnetic current. In the nonrelativistic limit,
\[
    \varepsilon^0=\mathcal O(v), \quad (p+p')^0\simeq2m_{\pi_d}, \quad q\cdot\varepsilon=\mathcal O(Q)
\]
The $C_1^A$ contribution is therefore suppressed relative to the $C_2^A$ contribution by $\sim \mathcal O(v\Lambda_d/Q)$. For halo DM producing the LZ event,
\[
    v \sim 10^{-3}, \quad \Lambda_d \sim \mathcal{O}(100)~\mathrm{MeV}, \quad Q \simeq 246~\mathrm{MeV},
\]
giving a $\sim 10^{-3}$ suppression. The scattering amplitude is consequently dominated by the $C_2^A$ term, corresponding to a
\begin{equation}
    J^\mu_{d, \rho\pi} \equiv -\frac{C_2^A(q^2)}{\Lambda_d}\partial_\nu (\pi^\dagger_d\overleftrightarrow{\partial^\mu} \rho^\nu_d) + \rm h.c.
\end{equation}
after Fourier transform.

For axion emission, the vector part of $J_{\rm PQ}^\mu$ is conserved and thus, gives no contribution after contraction with the axion momentum. Again, by Lorentz invariance and parity conservation, the axial transition elements are parameterized as
\begin{align}
     q_\mu\langle\pi_d(p)|J_{\rm PQ}^\mu|\rho_d(p',\varepsilon)\rangle
     ={}(q\cdot\varepsilon)
     \left[m_{\pi_d} C_1^a(q^2)
     +\frac{m_{\rho_d}^2-m_{\pi_d}^2}{\Lambda_d}C_2^a(q^2)
     +\frac{m_{a_d}^2}{\Lambda_d}C_3^a(q^2)\right]+ \cdots.
\end{align}
These coefficient $C_i^a$ include the hadronic matching and any anomalous pseudoscalar mixing. As
\[
    \frac{m_{\rho_d}^2-m_{\pi_d}^2}{\Lambda_d} \ll m_{\pi_d}, \quad \frac{m_{a_d}^2}{\Lambda_d} \ll m_{\pi_d},
\]
the contribution is dominated by the $C_1^a$ term. Define $C_{\rho\pi}\equiv C_{1}^a(m_{a_d}^2)$ relevant to the physical $\rho_d$ decay. This matches to a meson-level effective axion transition current
\begin{equation}
    J_{\rm PQ}^\mu=
     -m_{\pi_d}C_{\rho\pi}\,
     \pi_d^\dagger\rho_d^\mu+\mathrm{h.c.}.
\end{equation}
In the minimal theory, the axion phase enters through the light-quark mass, so the physical amplitude vanishes as $m_l\to0$. The light-quark mass $m_l$ depends on the unknown UV cutoff $\Lambda_{\rm UV}$ in Eq.~\eqref{eq:new_yukawa}. We assume $m_l$ is not too suppressed compared to $\Lambda_d$ when using $C_{\rho\pi}=\mathcal O(1)$.

\section{Cosmology}\label{app:cosmology}
\subsection{$\rho_d$ mesons}\label{app:rho_d}

We assume a primordial $h-l$ asymmetry sufficient to account for the observed DM abundance. Such an asymmetry could originate from CP-violating, out-of-equilibrium dynamics involving higher-dimensional $h-l$-violating interactions at high temperatures, analogous to baryogenesis in the visible sector~\cite{Sakharov:1967dj,Fukugita:1986hr,Huang:2024azp,Yoshimura:1978ex,Weinberg:1979bt,Affleck:1984fy,Pilaftsis:2003gt}. For example, in $SU(2)_d$, dimension-6 operators such as
\begin{equation}
    \frac{1}{\Lambda_{\rm UV}^2}
    \left(\epsilon_{ij}\chi_h^i\chi_l^j\right)
    \left(\chi^c_{h\,k}\chi_l^k\right)^\dagger
    \label{eq:asymmetry_operator}
\end{equation}
respect the gauge and PQ symmetries from Table~\ref{tab:charges} but carry non-trivial $(\Delta Q_h,\Delta Q_l)=(2,0)$ and hence $\Delta(Q_h-Q_l)=2$. Its interaction rate scales as $\Gamma\sim T^5/\Lambda_{\rm UV}^4$ at high temperature. Alternatively, an asymmetry generated in another sector could be transferred to the dark quarks through interactions that subsequently decouple~\cite{Zurek:2013wia}. Once the relevant flavor-violating interactions become ineffective, the accidental vector flavor symmetries $U(1)_h \times U(1)_l$ preserve the asymmetry, while dark strong annihilations deplete the symmetric component.

After confinement, the asymmetric heavy--light meson population contains both $\pi_d$ and $\rho_d$. The strong conversion $\rho_d\rho_d\leftrightarrow\pi_d\pi_d$ maintains the two species in relative chemical equilibrium until $T_{\rm f} \lesssim \Delta$, giving the yields ratio
\begin{equation}
    \frac{Y_{\rho_d}(T)}{Y_{\pi_d}(T)}
    =
    \frac{n_{\rho_d}(T)}{n_{\pi_d}(T)}
    \simeq 3e^{-\Delta/T}.
\end{equation}
Because both the conversion $\rho_d\rho_d\leftrightarrow\pi_d\pi_d$ and the decay $\rho_d\to\pi_d a_d$ preserve the total meson yield,
\begin{equation}
    Y_{\rm \pi_d}
        \simeq Y_{\pi_d}(T_{\rm f})+Y_{\rho_d}(T_{\rm f})
        \simeq
        4.38\times10^{-13}
        \left(\frac{\Omega_{\pi_d}h^2}{0.12}\right)
        \left(\frac{1\,\mathrm{TeV}}{m_{\pi_d}}\right),
        \label{eq:total_meson_yield}
\end{equation}
where $Y_{\pi_d}$ is today's $\pi_d$ yield. 

Taking $\langle\sigma v\rangle_{\rho_d\rho_d\to\pi_d\pi_d} \sim\Lambda_d^{-2}$, conversion freezes out when $n_{\rho_d}\langle\sigma v\rangle\sim H_{\rm RD}(T_{\rm f})$, one has
\begin{equation}
    T_{\rm f} \frac{3e^{-\Delta/T_{\rm f}}}{1+3e^{-\Delta/T_{\rm f}}}
        \sim \sqrt{\frac{45}{8\pi^2}} \frac{\sqrt{g_*(T_{\rm f})}}{g_{*s}(T_{\rm f})}\frac{\Lambda_d^2}{\MPL Y_{\pi_d}},
\end{equation}
where $\MPL=2.435\times10^{18}\,\mathrm{GeV}$, and $g_*$ and $g_{*s}$ count the effective energy and entropy degrees of freedom. For simplicity, we take the dark-sector and radiation temperatures to be roughly equal. At $T\sim\mathcal{O}(10)$ keV, $g_*=3.36$ and $g_{*s}=3.91$. For the benchmark $m_h \simeq m_{\pi_d}=750\,\mathrm{GeV}$, $\Delta=360\,\mathrm{keV}$, assuming $\kappa_{\rm hf}=3$, one has $\Lambda_d\simeq0.3\,\mathrm{GeV}$. With $\pi_d$ constituting the entire relic abundance, this gives
\begin{equation}\label{eq:decoupling_bm}
    T_{\rm f}\simeq41.7\,\mathrm{keV},
    \quad
    \left.\frac{Y_{\rho_d}}{Y_{\pi_d}}\right|_{T_{\rm f}}
    \simeq5\times10^{-4},
    \quad
    Y_{\rho_d}(T_{\rm f})\simeq3.1\times10^{-16}.
\end{equation}
Using $Y_{\pi_d}\propto m_{\pi_d}^{-1}$ and $\Lambda_d^2\simeq m_{\pi_d}\Delta/\kappa_{\rm hf}$, the decoupling temperature $T_{\rm f}$ is linearly sensitive to $\Delta$ and only logarithmically sensitive to $m_{\pi_d}$.

As illustrated in Eq.~\eqref{eq:axion_mass_bound}, PQ breaking at any $f_{a_d} \gg m_h$ ensures $m_{a_d}\ll\Delta$. Hence, the dark axion opens the decay $\rho_d\to\pi_d a_d$. The width from Eq.~\eqref{eq:rhopiawidth} is
\begin{equation}
    \Gamma_{\rho_d\to\pi_d a_d}
    \simeq
    \frac{|C_{\rho\pi}|^2\Delta^3}
         {216\pi f_{a_d}^2}.
\end{equation}
The characteristic decay temperature, defined by $H(T_{\rm dec})\sim\Gamma_{\rho_d\to\pi_d a_d}$, is
\begin{align}
    T_{\rm dec}
    &=
    \left(\frac{90}{\pi^2g_*(T_{\rm dec})}\right)^{1/4}
    \sqrt{\Gamma_{\rho_d\to\pi_d a_d}\MPL}
    \nonumber\\
    &\simeq
    0.53\,\mathrm{keV}\,
    |C_{\rho\pi}|
    \left(\frac{\Delta}{360\,\mathrm{keV}}\right)^{3/2}
    \left(\frac{10^9\,\mathrm{GeV}}{f_{a_d}}\right)
    \left(\frac{3.36}{g_*(T_{\rm dec})}\right)^{1/4}.
    \label{eq:rho_decay_temperature}
\end{align}
For $f_{a_d}\sim10^9\,\mathrm{GeV}$ and $|C_{\rho\pi}|\sim1$, conversion freezes out before the remaining excited states decay. These decays transfer the residual $\rho_d$ population into $\pi_d$, preserving the total meson abundance and eliminating the dangerous exothermic contribution to direct detection.

\subsection{Dark axion}

The initial displacement $f_{a_d}\theta_i$ produces a coherent axion population through misalignment~\cite{Preskill:1982cy,Abbott:1982af,Dine:1982ah,Baer:2026wre}. Oscillations begin at $T_{\rm osc}$, determined by
$3H(T_{\rm osc})\simeq m_{a_d}(T_{\rm osc})$, where $m_{a_d}(T)=\sqrt{\chi_d(T)}/f_{a_d}$. In the regime $\Lambda_d\ll T\ll m_h$, there is one active light dark-quark flavor. Neglecting logarithmic running, the leading high-temperature instanton result gives $\chi_d(T)\propto T^{-n}$, with $n=11(N-1)/3$~\cite{Boccaletti:2020mxu}. We approximate the interpolation to zero
temperature by
\begin{equation}
    \chi_d(T)\simeq\chi_d(0)
    \begin{cases}
        1, & T\leq\Lambda_d,\\[2pt]
        (\Lambda_d/T)^n, & \Lambda_d<T\ll m_h,
    \end{cases}
    \qquad
    n=\frac{11(N-1)}{3}.
    \label{eq:thermal_susceptibility}
\end{equation}
In the harmonic and sudden-onset approximation, this gives
\begin{align}
    \Omega_{a_d}^{\rm mis}h^2
    \simeq{}&
    \frac{2.55\times10^{-4}\,\theta_i^2}
         {(9.80\times10^6)^{2/(n+4)}}
    \left(
        \frac{f_{a_d}}{10^9\,\mathrm{GeV}}
    \right)^{\frac{n+6}{n+4}}
    \left(
        \frac{\chi_d(0)^{1/4}}{0.1\,\mathrm{GeV}}
    \right)^{\frac{2(n+2)}{n+4}}
    \left(
        \frac{0.5\,\mathrm{GeV}}{\Lambda_d}
    \right)^{\frac{n}{n+4}}
    \left(
        \frac{g_*(T_{\rm osc})}{100}
    \right)^{\frac{n+6}{2(n+4)}}
    \left(
        \frac{100}{g_{*s}(T_{\rm osc})}
    \right),
    \label{eq:axion_misalignment}
\end{align}
where $n=11(N-1)/3$. Using the parametric estimate $\chi_d(0)^{1/4}\lesssim\Lambda_d$ and $g_*(T_{\rm osc})\simeq g_{*s}(T_{\rm osc})\simeq100$, we obtain
\begin{equation}
    \Omega_{a_d}^{\rm mis}h^2
    \lesssim
    \frac{6.38\times10^{-3}\,\theta_i^2}
         {(2.45\times10^8)^{2/(n+4)}}
    \left(
        \frac{f_{a_d}}{10^9\,\mathrm{GeV}}
    \right)^{\frac{n+6}{n+4}}
    \left(
        \frac{\Lambda_d}{0.5\,\mathrm{GeV}}
    \right).
    \label{eq:misalignment_bound}
\end{equation}
With $f_{a_d}=10^9\,\mathrm{GeV}$, $\chi_d(0)^{1/4}=\Lambda_d=0.3\,\mathrm{GeV}$, $\theta_i=1$, and $g_*=g_{*s}=100$, we obtain
\begin{align}
    \Omega_{a_d}^{\rm mis}h^2&\simeq
    \begin{cases}
        2.48\times10^{-5}, & SU(2)_d,\\
        1.27\times10^{-4}, & SU(3)_d,\\
        2.91\times10^{-4}, & SU(4)_d,
    \end{cases}
\end{align}
and remain at $\mathcal{O}(10^{-4})$ even for very large $n$. Thus, the misalignment component accounts for only a tiny fraction of the observed dark-matter density within the region of interest.

Late $\rho_d$ decays contribute to the effective number of neutrino species $N_{\rm eff}$. Once $\rho_d\to\pi_da_d$ is thermally open, each decay produces an axion with energy $E_{a_d}\simeq\Delta$. In the sudden-decay approximation, the injected energy density satisfies
\begin{equation}
    \left.\frac{\rho_{a_d}^{\rm dec}}{s}
    \right|_{T_{\rm dec}}
    \simeq
    \Delta\,Y_{\rho_d}(T_{\rm f}).
    \label{eq:axion_decay_injection}
\end{equation}
As $m_{a_d} \ll \Delta$, these axions remain relativistic at recombination. Their contribution is
\begin{align}
    \Delta N_{\rm eff}^{\rm dec}
    &=
    \frac{8}{7}\left(\frac{11}{4}\right)^{4/3}
    \frac{\rho_{a_d}^{\rm dec}}{\rho_\gamma}
    \simeq
    \frac{18\,\Delta}
         {g_{*s}^{1/3}(T_{\rm dec})T_{\rm dec}}\,
    Y_{\rho_d}(T_{\rm f})
    \nonumber\\
    &\simeq
    1.3\times10^{-13}
    \left(\frac{Y_{\rho_d}(T_{\rm f})}{10^{-16}}\right)
    \left(\frac{\Delta}{360\,\mathrm{keV}}\right)
    \left(\frac{3.2\,\mathrm{keV}}{T_{\rm dec}}\right)
    \left(\frac{3.91}{g_{*s}(T_{\rm dec})}\right)^{1/3}.
    \label{eq:neff_decay}
\end{align}
The small residual excited-state yield makes this dark-radiation contribution completely negligible compared with the CMB constraint at 95\% confidence~\cite{AtacamaCosmologyTelescope:2025nti}
\begin{equation}
    \Delta N_{\rm eff} \lesssim 0.17.
\end{equation}

\end{document}